\documentstyle[12pt,epsfig]{article}
\def\v1{\vspace{1cm}}
\def\be{\begin{equation}}
\def\ee{\end{equation}}
\def\bc{\begin{center}}
\def\ec{\end{center}}
\def\ik{\partial}
\def\vh{\varphi}

\newcommand{\bea}{\begin{eqnarray}}
\newcommand{\eea}{\end{eqnarray}}

\begin{document}
\setlength{\baselineskip}{5mm}
\hfill
\parbox{4cm}{\mbox{JINR-E2-2001-52}}
\\[1cm]
\noindent{\large\bf
Cosmological Creation of Vector Bosons and Fermions
}\vspace{4mm}

\noindent{
D.~Blaschke, V.~Pervushin, D.~Proskurin, S.~Vinitsky, and A.~Gusev
}\vspace{1mm}

\noindent{\small
Joint Institute for Nuclear Research, 141980 Dubna, Russia
}\vspace{4mm}

\begin{abstract}{
The cosmological creation of primordial vector bosons
and fermions is described in the Standard Model of strong and electro-weak
interactions given in a space-time with
the relative standard of measurement of geometric intervals.
Using the reparametrization - invariant perturbation theory and
the holomorphic representation of quantized fields we derive
equations for the Bogoliubov coefficients and distribution functions of
created particles.
The main result is the intensive cosmological creation of
longitudinal Z and W bosons (due to their mass singularity)
by the universe in the rigid state.
We introduce the hypothesis that the decay of the primordially created
vector bosons
is the origin of the Cosmic Microwave Background radiation.
}
\end{abstract}

\section{Introduction}

It was the main achievement of the Hot Big Bang Cosmology
to predict the Cosmic Microwave Background (CMB) radiation
as one of the observable relics witnessing the thermal history
of the expanding universe.
The discovery of the CMB has in turn disfavored alternative cosmologies 
like the conformal-invariant Hoyle-Narlikar type cosmology~\cite{N}.
The latter has recently been generalized and successfully applied
to a description of luminosity distances for high redshift supernovae
~\cite{ps} without the need for a $\Lambda$ term thus solving one of the 
major problems of the Standard Cosmology~\cite{12}. 
In this Conformal Cosmology, the Hubble law is explained by the cosmic 
evolution of elementary particles masses. The question arises about the
origin of the CMB in the context of the Conformal Cosmology.
In the present work we consider the simplest scenario where
the origin of the CMB is the primordial creation of
longitudinal vector bosons which have a singular behavior of the integral
of motion in the vicinity of the cosmological singularity~\cite{sf,hp}.

In this paper, we give a systematic description of the cosmological
creation of massive vector bosons and fermions in the Standard Model of
strong and electroweak interactions.
Our consideration of this problem differs from other calculations
~\cite{bt,par,gmm,zel}
by i) the conformal symmetry~\cite{plb}, ii) the reparametrization -
invariance~\cite{grg,pp}, and iii) the holomorphic representation of
quantized fields~\cite{ps1}.

The conformal symmetry was introduced in the theory
of gravitation by Weyl in 1918~\cite{we}, based on the fact that we can 
measure only a ratio of two intervals. 
However, the original Weyl theory of a {\it vector} field had the defect 
of an ambiguous
physical time as marked by Einstein in his comments to Weyls work.
The conformal invariant theory of gravitation with an unambiguous
physical time has been considered in Ref.~\cite{plb} on the basis of
the conformal invariant theory of a massless {\it scalar} (dilaton)
field~\cite{pct} with a negative sign.

The  dilaton theory of gravitation is mathematically equivalent
to the Einstein theory, and all solutions
of the one theory  can be constructed from those of the other by
conformal transformations. In particular, the homogeneous approximation
in Einstein's theory corresponds to the lowest order of the
reparametrization - invariant perturbation theory in the flat space-time
in the conformal invariant theory, where the homogeneous dilaton field 
scales all masses including the Planck mass. The corresponding
Conformal Cosmology~\cite{12}
describes the cosmic evolution of all masses~\cite{N} with respect to
the observable conformal time
instead of the cosmic evolution of the scale factor in the
Standard Cosmology.

The considered perturbation theory keeps the main symmetry of all metric
gravitation theories-
the invariance with respect to reparametrizations of the "coordinate
time". Just this invariance leads to
the energy constraint that connects the total energy of all fields
in the universe with the
energy of the dilaton and converts it into
the evolution parameter of the history of the universe.

The content of the paper is the following.
In Section 2, we formulate the conformal - invariant version of the
Standard Model (SM) unified with the dilaton version of Einsteins
General Relativity (GR) theory.
Section 3 is devoted to the reparametrization - invariant perturbation
theory as the basis of the Conformal Cosmology.
In Section 4, we derive equations of
the cosmological creation of vector bosons and fermions.
In Section 5, we solve these equations
for the early universe in the rigid state.
We summarize the results of this work in the Resum\`e of Section 6 and 
give our conclusions in the final Section 7.

\section{Conformal General Relativity}

\subsection{Conformal - invariant unified theory}

We consider a version of the conformal-invariant unified theory of
gravitational, electroweak and strong interactions as the Standard Model
where the dimensional parameter in the Higgs potential is replaced by
the dilaton field $w$ the dynamics of which is described by the negative 
Penrose-Chernikov-Tagirov action. The corresponding action  takes the form
\be \label{Cw}
W_{\rm C}=\int d^4x \sqrt{-g}\left[
{\cal L}_{\Phi,w}+
{\cal L}_{g}+{\cal L}_{l}+
{\cal L}_{l\Phi}\right],
\ee
where
\be \label{LP}
{\cal L}_{\Phi,w}=\frac{|\Phi|^2-w^2}{6}R-\partial_{\mu}w\partial^{\mu}w
+D^{-}_{\mu}\Phi(D^{\mu,-}\Phi)^*
-\lambda\left( |\Phi|^2 - y^{2}_{h}w^{2} \right)^{2}
\ee
is the Lagrangian of dilaton and Higgs fields with 
$ D_{\mu}^{-}\Phi=\left(\partial_{\mu} -
\imath g \frac{\tau_{a}}{2}A_{\mu}-
\frac{\imath}{2}g^{\prime}B_{\mu}\right)\Phi$, and 
\be \label{Ph}
\Phi = \left(
 \begin{array}{rr}
 \Phi_{+} \\
 \Phi_{0}\end{array}\right);~~(|\Phi|^{2} = 
\Phi_{+}\Phi_{-} + \Phi_{0}\bar{\Phi}_{0})~.
\ee
The Lagrangians of the gauge fields is
\be \label{LG}
{\cal L}_{g}=-\frac{1}{4}\left(\partial_{\mu}A^{a}_{\nu} -
\partial_{\nu}A^{a}_{\mu}+g\varepsilon_{abc}A^{b}_{\mu}A^{c}_{\nu}
\right)^{2}-\frac{1}{4}\left(\partial_{\mu}B_{\nu}
- \partial_{\nu} B_{\mu} \right)^{2}~,
\ee
and the Lagrangian of leptons
\be \label{L}
L = \left(
 \begin{array}{rr}
 \nu_{e,L} \\
  e_L \end{array}\right);~~e_R,\nu_{e,R},
\ee
is given by
\be \label{LL}
{\cal L}_{l}=\bar{L}\imath \gamma^{\mu}\left(D^F -
\imath g \frac{\tau_{a}}{2}A_{\mu}+ \frac{\imath}{2}g^{\prime}B_{\mu}\right)L
+ \bar{e}_{R}\imath \gamma^{\mu}\left(D^F_{\mu} +
\imath g^{\prime}B_{\mu}\right)e_{R} +
\bar{\nu}_{R}\imath \gamma^{\mu}\partial_{\mu}\nu_R~,
\ee
where $D^F$ is the Fock derivative. The Lagrangian describing the masses of 
the leptons is
\be \label{LVh}
{\cal L}_{l\vh}= - y_{e}\left(e_{R}\Phi^{+}L + \bar{L}\Phi e_{R}\right)
- y_{\nu}\left(\bar{\nu}_{R}\Phi^{+}_{C}L + \bar{L}\Phi_{C}\nu_{R} \right),~~
\Phi_{C} = \imath \tau_{2}\Phi^{+}_{C} =
\left(\begin{array}{rr}
 \bar{\Phi}_{0} ~\\
 - \Phi_{-}\end{array}\right)~,
\ee
which includes the possibility of a finite neutrino mass, $y_f$ are 
dimensionless parameters.

This theory is invariant with respect to conformal transformations,
and it is given in the Weyl space of similarity with the relative
standard of the measurement of intervals given by the ratio
of two intervals~\cite{we}
\be \label{ratio}
r= \frac{ds^2}{ds^2_{\rm scale}}~.
\ee
The space of similarity is the manifold of Riemannian spaces connected
by the conformal transformations.

The ratio~(\ref{ratio}) in the geometry of similarity
depends on nine components of the metric tensor,
the relative standard measurement of intervals allows us to remove
the scale variable $|{}^{(3)}g|$ from the metric tensor.
Therefore, we use the conformal-invariant Lichnerowicz
variables~\cite{L} and the 
measurable conformal-invariant space-time interval
\be
\label{dse}
  (ds^L)^2=g^L_{\mu\nu}dx^\mu dx^\nu~~~~~~~~~
,~~~
g^L_{\mu\nu}=g_{\mu\nu}|{}^{(3)}g|^{-1/3},~~~|{}^{(3)}g^L|=1~,
\ee
with the notation $f_n^L=f_n|{}^{(3)}g|^{n/6}$, where $(n)$ is the conformal 
weight being equal to $(1,3/2,0,-2)$
for the scalar, spinor, vector, and tensor field respectively.
The Lichnerowicz interval depends on
nine components of the metric $g^L_{\mu\nu}$.

The formulation of GR in terms of the Lichnerowicz variables reveals that
the equivalence principle is violated in
a direct unification of GR and SM where the gravitational coupling constant
and inertial masses of particles are formed by different fields, i.e. when 
the scale component of metric and the Higgs field are treated as independent 
scalar fields. A possible solution to this problem is given in the following 
subsection.

\subsection{Conformal Higgs Mechanism}

In the contrast to the direct unification of GR and SM,
the principle of equivalence of inertial and gravitational masses
is incorporated into the conformal invariant Higgs mechanism through the
dilaton-Higgs mixing~\cite{p1}
\be \label{p1}
w=\phi~ch\chi,~~~~~~\Phi_i=\phi~{\bf n}_i sh\chi,~~~ ({\bf n}{\bf n}^{+}=1).
\ee
The modulus of dilaton-Higgs mixing $\phi$ simultaneously forms
the gravitational coupling constant and inertial masses of particles.

After applying the transformation~(\ref{p1}) the Lagrangian in
the action~(\ref{Cw}) takes the form
\be \label{LCGR}
{\cal L}_{C}= - \frac{\phi^{2}}{6}R + \frac{\phi}{\sqrt{-g}}\partial_{\mu}
\left(\sqrt{-g} \partial^{\nu}\phi\right) +
\phi^{2}\partial_{\mu}\chi\partial^{\mu}\chi
+ \bar{{\cal L}}_{SM},
\ee
where $\bar{{\cal L}}_{SM}$ is the SM Lagrangian
with the Higgs potential
\be \label{Higgs}
{\cal L}_{Higgs}= -\lambda\phi^{4}\left(sh^2\chi-y^{2}_{h} ch^2\chi\right)^{2}.
\ee
The extrema of this potential with respect to $\chi$ can be found from the 
condition
\be \label{Var}
\frac{\partial {\cal L}_{Higgs}}{\partial \chi}
= -4\lambda\phi^{4}(1-y_{h}^{2})sh\chi ch\chi
\left(sh^2\chi (1 - y_{h})^{2} - y_{h}^{2} \right)= 0~,
\ee
which leads to two solutions
\be \label{Sol}
\chi_1 = 0,~~~~|sh \chi_2| = \frac{y_h}{\sqrt{1 - y^{2}_h}}.
\ee
The last solution corresponds to the spontaneous SU(2) symmetry breaking
in the SM. The problem is to show that the present day value of the modulus
of the dilaton-Higgs mixing $\phi(t_0,x)=\vh_0$ far from heavy masses is equal to
the Newton constant
\be \label{crgr}
\vh_0 = M_{\rm Planck}\sqrt{\frac{3}{8\pi}}~.
\ee
This means, that
\be \label{yh}
y_h=\frac{m_h}{\vh_0}\sim 10^{-17}~.
\ee
In terms of the notations $\vh_0\chi = H $ and  $\vh_0 y_h = \bar{M}_h$
we obtain in the limit of the infinite Planck mass
the renormalizable version of the SM with the Higgs potential
\be \label{yh1}
-\lambda(H^2 - \bar{M}^{2}_h)^{2} + O(\frac{1}{M_{\rm Planck}}).
\ee
On the other hand, if $\lambda = 0$, then the SU(2) breaking solution is
$\chi =$ const, and  we obtain the Higgs free SM version~\cite{plb}.

In the following, we consider the problem of cosmological particle creation
in the framework of the dilaton theory~(\ref{Cw}).


\section{The Energy Constrained Perturbation Theory}

\subsection{Conformal Cosmology}

The cosmological applications of conformal gravity are based on
the perturbation theory~\cite{ps1} that begins from
the homogeneous approximation for the dilaton and metrics
\be \label{crin1}
 \phi^L(t,x) = \vh(t),~~~~[g^{00}_L(t,x)]^{-1/2}=N_0(t),
~~~~~~~ g^L_{ij}=\delta_{ij}+h_{ij}~,
\ee
where the conformal - invariant interval reads
\be \label{crin0}
ds^2_0=d\eta^2 - dx_i^2~,~~~~~~~~~~~~~~~~    d\eta = N_0(t)dt~.
\ee
We keep only independent local field variables which are  determined by
a complete set of initial values. All nonphysical variables (for which the 
initial values depend on other data) are excluded by the local constraint.
In particular, the local part of the dilaton is excluded by the
local energy constraint, i.e. the equation for the metric
component $g_L^{00}$, and is converted into the Newtonian interaction
potential. 
The first step of the perturbation theory is to consider the independent 
"free" fields in the linear approximation of their equations of motion.
The global part of the dilaton $\vh(t)$ should be considered as
independent variable as its two initial values (the field and its momentum) 
cannot be determined by the one energy constraint only~\cite{pp}.
The latter is the consequence of the invariance of the theory with respect to
reparametrizations of the "coordinate time" $t\rightarrow \bar t=\bar t(t)$.

The substitution of the ansatz~(\ref{crin0}) into the
action~(\ref{Cw}) leads to the action of free fields in terms of
physical variables~\cite{hp,ps1}
\be \label{gradl}
W_0=\int\limits_{t_1 }^{t_2 }dt \left[ \vh \frac{d}{dt}\frac{\dot
\vh}{N_0}V_0 + N_{0}L_0\right]~,
\ee where $V_0$ is a finite
spatial volume and
\be \label{H0l}
L_0=\frac{1}{2}
\int\limits_{V_0 } d^3x N_0
\left({\cal L}_{\rm \chi} + {\cal L}^{\bot}_{\rm vec}+ {\cal
L}^{||}_{\rm vec}+{\cal L}_{\rm rad}+{\cal L}_{\rm s}+
{\cal L}_{\rm h}\right),
\ee
is the total Lagrangian of free fields.
In particular,
\be \label{Hig}
{\cal L}_{\rm \chi}=\vh^{2}\left(\frac{\dot{\chi}_i^{2}}{N^{2}_{0}}+
\chi_i\left[ \vec{\partial}^2-4\lambda(y_h\vh)^2\right] \chi_i\right)
\ee
is the Lagrangian of the Higgs field and 
\begin{eqnarray}
{\cal L}_{\rm vec}^{\bot} &=& {1\over
2}\left(\frac{\dot{v}_i^{{\bot}2}}{N^{2}_{0}}+
v_i^{\bot}\left[ \vec{\partial}^2-(y_v\vh)^2\right] v_i^{\bot}\right)~,
\nonumber\\
{\cal L}_{\rm vec}^{||} &=& -{1\over 2}(y_v\vh)^2
\left(\frac{\dot{v}_i^{||}}{N_0}{1\over
\left[\vec{\partial}^2-(y_v\vh)^2\right]}
\frac{\dot{v}_i^{||}}{N_0}+v_i^{||2}\right)
\end{eqnarray}
are Lagrangians of the transversal (${\cal L}_{\rm vec}^{\bot}$)
and longitudinal (${\cal L}_{\rm vec}^{||}$) components of the
W- and Z- bosons~\cite{sf,hp}.
The Lagrangian of the fermionic spinor fields is given by
\be \label{hml}
{\cal L}_{s}= \bar
\psi\left(-y_s\vh- i\frac{\gamma_0}{N_0}\ik_0 +i\gamma_j\ik_j\right)\psi~,
\ee
where the r\^ole of the masses
is played by the homogeneous dilaton field $\vh $ multiplied by
dimensionless constants $y_{v,s}$. ${\cal L}_{rad}$ is the
Lagrangian of massless fields (photons $\gamma$, neutrinos $\nu$)
with $y_{\gamma}=y_{\nu} =0$, and
\be \label{hol}
{\cal L}_{h}=\frac{\vh^2}{24}\left(\frac{\dot h^2}{N^{2}_{0}}-
(\ik_ih)^2\right)~
\ee
is the Lagrangian of gravitons as weak transverse excitations of spatial
metric for which $h_{ii}=0;~\ik_jh_{ji}=0~$. 
The last two equations follow from the unit determinant
of the three-dimensional metric~(\ref{dse}) and from the momentum
constraint~\cite{ps1}.

To find the evolution of all fields with respect to the
proper time $\eta$, we use the Hamiltonian form of
the action~(\ref{Cw}) in the approximation~(\ref{crin0})
\be \label{grad}
W_0^E=\int\limits_{t_1 }^{t_2 }dt
\left( \left[\int\limits_{V_0 }
d^3x \sum\limits_{f}p_f\dot f \right]
-\dot \vh P_{\vh}-N_{0}\left[-\frac{P_{\vh}^2}{4V_0}
+H(\vh,f,p_f)\right]\right)~,
\ee
where the Hamiltonian $H(\vh,f,p_f)$ is a sum of the Hamiltonians of free 
fields, with $f$ and denoting the field of particle species $f$ and  $p_f$ 
its conjugate momentum.

The variation of the action with respect to
the homogeneous lapse-function $N_0$ yields the energy constraint
\be
\label{d56}
\frac{\delta W_{0}}{\delta N_0}=0~
\Rightarrow\,
\vh'^2=\frac{\bar H(\vh)}{V_0}:=\rho(\vh)~,
\ee
where the prime denotes the derivative with respect to the
conformal time $\eta$. 
The fact that the dilaton, which scales all the elementary particle masses, 
has a nonzero derivative entails the cosmic evolution of the size of atoms 
similar to the Hoyle-Narlikar cosmology~\cite{N,12,plb}.
The cosmic evolution of the dilaton leads to a rescaling of the
energy levels of  atoms and to the observed redshift $z$ of atomic line spectra
$$
z+1=\vh(\eta_0)/\vh(\eta_0-d/c)\simeq 1+
(d/c)H_0~,
$$
of a star at the distance $d$ from the Earth
where $H_0=(\log\vh)'|_{\eta_0}$ is the conformal cosmology definition of the 
Hubble parameter.
With these definitions, the present-day value of the dilaton 
$\vh_0=\vh(\eta_0)$ can be determined 
from the present-day value of density parameter $\Omega_0=\rho(\vh_0)/\rho_c$,
where $\rho_c=3 M_{\rm Planck}^2H_0^2/(8 \pi)$,
\be \label{cg1}
\vh_0=\Omega^{-1/2}_{0} M_{\rm Planck}\sqrt{\frac{3}{8\pi}}~.
\ee

Thus, the energy constrained theory 
obeys the Freedmann equation of the cosmological evolution
\be \label{gt1}
\eta(\vh_0,\vh_I)=\pm \int\limits_{\vh_I }^{\vh_0 }
\frac{d\vh}{\sqrt{\rho(\vh)}}
\ee
that connects the geometric interval with the dynamics of the dilaton as
evolution parameter.
In accordance with the Dirac quantization of relativistic
constrained systems the geometric time~(\ref{gt1}) is always 
positive~\cite{pp}.

\subsection {Holomorphic Representation of Quantized Fields}

In order to determine the observational energy density~(\ref{d56}) we 
have to diagonalize the Hamiltonian
\be \label{h} 
{\hat H} = \sum_{\varsigma,J}\omega^J_{\varsigma}(\vh){\hat N}_{\varsigma}^J~,
\ee
where $\varsigma = k, f, \sigma$ stands for momenta,
species, and spins, respectively. 
${\hat N}_\varsigma^J$ are the number operators
for bosons ($J=B$) and fermions ($J=F$), given by
\bea
{{\hat N}^B}_{\varsigma} &=&
\frac{1}{2}(a_{\varsigma}^{+}a_{\varsigma} +
a_{\varsigma}a_{\varsigma}^{+})~,\\
{\hat N}^{(J=F)}_{\varsigma} &=&
(a_{\varsigma}^{+}a_{\varsigma} -
d_{\varsigma}d_{\varsigma}^{+})~.
\eea
We introduce the representation of particles as holomorphic field variables
\be
\label{grep}
   f(t,\vec x)=\sum\limits_{k}^{ }
\frac{C_f(\vh)\exp(ik_ix_i)}{V_0^{3/2}\sqrt{\omega_f(\vh,k)}} \bar f_J(t,k)~,
\ee
where
\bea
\label{holb}
\bar f_B(t,k)&=&\frac{1}{\sqrt{2}}\sum_{\sigma}
\left( a_{\sigma}^+(-k,t)\epsilon_{\sigma}(-k)+
a_{\sigma}(k,t)\epsilon_{\sigma}(k)\right)~,\\
\label{holf}
\bar f_F(t,k)&=&\sum_{\sigma}
\left( d_{\sigma}^+(-k,t)v^*(-k)+ a_{\sigma} (k,t)u_{\sigma}(k) \right)~,
\eea
and $\omega_{f}(\vh,k)=\sqrt{k^2+y_f^2\vh^2}$ are the one-particle
energies for particle species $f=h,\gamma,v,s,\chi$
with the dimensionless mass parameters $y_f$ and
$$
C_h(\vh) = \frac{\sqrt{12}}{\vh},~~~~
C_{\gamma}(\vh)=C_{\rm s}(\vh)=1,~~~~ C^{\bot}_{v}=1,~~~~
C^{||}_{v}=\frac{\omega_{v}}{y_{v}\vh},~~~C_{\chi}(\vh) = \frac{\sqrt{2}}{\vh}.
$$
The  coefficients ($C_h,C^{||}_{v}, C_{\chi}$) exhibit the mass
singularity of gravitons,  longitudinal components of massive vector
fields~\cite{hp}, and the Higgs fields. 
These mass singularities lead to the intensive cosmological
creation of the corresponding particles that follows from the first terms of 
the action~(\ref{grad}) when represented in terms of holomorphic variables
\bea
\label{actb}
\left[\int\limits_{V_0 }d^3x \sum\limits_{f}p_f\dot f \right]_B
&=&\sum_{\varsigma}
\frac{\imath}{2}( a^{+}_{\varsigma} {\dot a}_{\varsigma}
-a_{\varsigma}{\dot a}^{+}_{\varsigma} ) -
\sum_{\varsigma}
\frac{\imath}{2}(a^{+}_{\varsigma}a^{+}_{\varsigma} -
a_{\varsigma}a_{\varsigma})
{\dot \Delta}_{\varsigma}(\vh),
\\
\label{actf}
\left[\int\limits_{V_0 }d^3x \sum\limits_{f}p_f\dot f \right]_F
&=&\sum_{\varsigma}\imath( d_{\varsigma} {\dot d}_{\varsigma}^+
+a^+_{\varsigma}{\dot a}_{\varsigma} ) +
\sum_{\varsigma}
\imath(a^{+}_{\varsigma}d^{+}_{\varsigma} -
d_{\varsigma}a_{\varsigma})
{\dot \Delta}_{\varsigma}(\vh)~.
\eea
$\Delta_{\varsigma}(\vh)$ are terms responsible for
particle creation,
\bea 
\label{th} 
\Delta_{h}(\vh) &=& \ln(\vh) - \ln(\vh_{I})~,
\\ 
\label{ttv} 
\Delta^{{\bot}}_{v}(\vh) &=&
\ln(\sqrt{\omega_{v}}) -  \ln(\sqrt{\omega_{I}})~,
\\ 
\label{tv}
\Delta^{||}_{v}(\vh) &=&\Delta_{h}(\vh) - \Delta^{{\bot}}_{v}(\vh)~,
\\
\label{tc}
\Delta_{\chi}(\vh) &=&\Delta_{h}(\vh) + \Delta^{{\bot}}_{v}(\vh)~,
\\
\label{ts} 
\Delta_{s}(\vh) &=& \sigma \left[
\arctan\left(\frac{y_s \vh}{k}\right)
- \arctan\left(\frac{y_s \vh_I}{k}\right)\right]
~,~~~\sigma = \pm \frac{1}{2}~,
 \eea
where $\vh_I$ and $\omega_I$ are initial data.

\section {Cosmological Pair Creation}

\subsection {Bogoliubov Quasiparticles}

The local equations of motion for the system (\ref{grad}) can be written 
as~\cite{ps1}
 \be
 -\imath \frac{d}{d\eta}\chi_{\varsigma} = -\imath \chi'_{\varsigma} =
{\hat H}_{\varsigma}(\vh)\chi_{\varsigma}~,
 \ee
 where
 \be
\hat H_{\varsigma} = \left(
 \begin{array}{ll} \omega_{\varsigma}, & - \imath \Delta'_{\varsigma}\\ \\
 - \imath \Delta'_{\varsigma}, & - \omega_{\varsigma} \end{array}\right),
~~~~~~~
 \chi^{(B)}_{\varsigma} = \left(
 \begin{array}{rr}
 a^{+}_{\varsigma} \\
 a_{\varsigma} \end{array}\right),~~~~
 \chi^{(F)}_{\varsigma} = \left(
 \begin{array}{rr}
 d^{+}_{\varsigma} \\
 a_{\varsigma} \end{array}\right),~~
\ee
for bosons $(B)$ and fermions $(F)$, respectively.
Exact solutions of these equations of motion were obtained
by their diagonalization with the Bogoliubov transformations \cite{ps1}
\be
 \chi_{\varsigma}=\hat{O}_{\varsigma}\psi_{\varsigma}
 , ~~\Rightarrow ~~
 -\imath \psi'_{\varsigma}=\left[\imath\hat O{}_{\varsigma}^{-1}
 \hat O'_{\varsigma} +
 \hat O{}_{\varsigma}^{-1}\hat H_{\varsigma}{}\hat O_{\varsigma}\right]
\psi_{\varsigma}\equiv \bar H{}_{\varsigma}\psi_{\varsigma}~,
 \ee
 where $\bar H_{\varsigma}$  is required to be diagonal
 \be
 \bar H{}_{\varsigma} = \left(
 \begin{array}{rr} \bar{\omega}_{\varsigma},&0\\ \\ 0,
 &-\bar{\omega}_{\varsigma}\end{array}\right);
 ~~~~ \det \hat O_{\varsigma} = 1; ~~~~
 \psi^{(F)}_{\varsigma} =
 \left( \begin{array}{rr} b_{\varsigma}^{+} \\ c_{\varsigma}
 \end{array}\right),~~
 \psi^{(B)}_{\varsigma} =
 \left( \begin{array}{rr}
b_{\varsigma}^{+} \\ b_{\varsigma} \end{array}\right)~.
\ee
The Bogoliubov parametrizations for the coefficients are
\bea
b_{\varsigma}^{+}&=&\cos(r_{\varsigma})e^{-i \theta_{\varsigma}}
d_{\varsigma}^{+}
+\imath \sin(r_{\varsigma})e^{i \theta_{\varsigma}}a_{\varsigma}~,
\nonumber\\
c_{\varsigma}&=&\cos(r_{\varsigma})e^{i \theta_{\varsigma}}a_{\varsigma} +
\imath\sin(r_{\varsigma})e^{-i \theta_{\varsigma}}d_{\varsigma}^{+}~,
\eea
and the Hermitian conjugate for fermions, and
\bea
b_{\varsigma}^{+}&=&\cosh(r_{\varsigma})e^{-i \theta_{\varsigma}}
a_{\varsigma}^{+}
        -\imath \sin(r_{\varsigma})e^{ i \theta_{\varsigma}}a_{\varsigma}~,
\nonumber\\
b_{\varsigma}   &=&\cosh(r_{\varsigma})e^{ i \theta_{\varsigma}}a_{\varsigma}
       + \imath\sinh(r_{\varsigma})e^{-i \theta_{\varsigma}}a_{\varsigma}^{+}
\eea
for bosons.
For each ${\varsigma}$ we get two equations for the two unknown functions
$r_{\varsigma}, \theta_{\varsigma}$ 
\bea
[\omega_{\varsigma} - \theta'_{\varsigma}] \sinh(2r_{\varsigma}) &=&
\Delta'_{\varsigma}\cos(2\theta_{\varsigma})\cosh(2r_{\varsigma})~,
\nonumber\\
\label{main}
r'_{\varsigma} &=& - \Delta'_{\varsigma}\sin(2\theta_{\varsigma})
\eea
for bosons, and
\bea
[\omega_{\varsigma} - \theta'_{\varsigma}] \sin(2r_{\varsigma}) &=&
\Delta'_{\varsigma}\cos(2\theta_{\varsigma})\cos(2r_{\varsigma})~,
\nonumber\\
\label{mains}
r'_{\varsigma} &=& - \Delta'_{\varsigma}\sin(2\theta_{\varsigma})
\eea
for fermions.

The diagonalization procedure has two immediate consequences: 
(i) integrals of motion of the type of numbers of the Bogoliubov quasiparticles
are obtained and (ii) the possibility to choose the initial states as 
corresponding to the vacuum of the Bogoliubov quasiparticles (the well-known
{\it squeezed vacuum}~\cite{ps1})
\be \label{vacuum}
b_{\varsigma}|0>_{\rm sq} = 0~.
\ee
In the case of the vacuum initial data~(\ref{vacuum}),
the Bogoliubov equations~(\ref{main}) are closed by Eq.~(\ref{d56})
for evolution of the universe
\be \label{vdens}
\vh'^{2} = \frac{1}{V_0} \sum_{\varsigma} \omega_{\varsigma}(\vh)
{}_{\rm sq}<0|\hat N_{\varsigma}|0>_{\rm sq}
= \rho(\vh)~.
\ee
The number of particles created during the time $\eta$ is
\be \label{Nb}
{\cal N}_{\varsigma}^{(B)}(\eta) =
{}_{\rm sq}<0|\hat N^{(B)}_{\varsigma}|0>_{\rm sq}-
\frac{1}{2}=\sinh^2 r_{\varsigma}(\eta)
\ee
for bosons, and
\be \label{Nf}
{\cal N}_{\varsigma}^{(F)}(\eta) =
{}_{\rm sq}<0|\hat N^{(F)}_{\varsigma}|0>_{\rm sq}+
\frac{1}{2}=\sin^2 r_{\varsigma}(\eta)
\ee
for fermions respectively.

\subsection{The redshift representation}

To compare with the present-day cosmological data
${\rho_0}$ and $\vh_0$, it is useful
to rewrite the Bogoliubov equations~(\ref{main}) in terms of
the redshift $z$ and the density parameter $\Omega(z)$ defined as
\be\label{red}
z+1=\frac{\vh_0}{\vh},~~~~~~~~~~~~~~~~~~~~~
\Omega(z)=\frac{\rho}{\rho_c}~.
\ee
Then Eq.~(\ref{vdens}) for the evolution of the universe takes the form
(we take $\Omega_0=1$ from now on)
\be\label{vd}
H(z):=\frac{\vh'}{\vh}=-\frac{z'}{z+1}= (z+1) \sqrt{\Omega(z)} H_0~,
\ee
where $H_0=\vh'_0/\vh_0$ is a value of the  present-day Hubble parameter.
All equations can be rewritten in the terms of z-factor, in particular,
the Bogoliubov equations~(\ref{main})  for bosons become
$$
\sinh(2r_{\varsigma})
[\frac{\omega_{\varsigma}}{z'} -
\frac{d}{dz}\theta_{\varsigma}] =
\cos(2\theta_{\varsigma})\cosh(2r_{\varsigma})
\frac{d}{dz}\Delta_{\varsigma},
$$
\be \label{mainz}
\frac{d}{dz}r_{\varsigma}= -\sin(2\theta_{\varsigma})
 \frac{d}{dz}\Delta_{\varsigma}~,
\ee
where $z'$ is determined by eq.~(\ref{vd}).

\subsection{The vacuum initial values}

In the case of the early universe,
we have a large current Hubble parameter~(\ref{vd}) or small one-particle
energies
$\Delta' \ll \omega$. To analyze the Bogoliubov equations~(\ref{main})
in this case, we change variables
$(r,\theta\rightarrow C,{\cal N})$
\bea 
\label{sub}
\cos(2\theta_{\varsigma})\sinh(2r_{\varsigma})&=&C_{\varsigma}~,\nonumber\\
\sinh(2r_{\varsigma})&=&\sqrt{{\cal N}({\cal N}+1)}~.
\eea
Then, equations~(\ref{main}) take the form
\bea \label{3}
{\cal N}'_{\varsigma}&=&
\left( \frac{\Delta'_{\varsigma}}{2\omega_{\varsigma}} \right)C_{\varsigma}'
~,\nonumber\\
{\cal N}'_{\varsigma}&=&
-\Delta'_{\varsigma}\sqrt{ 4 {\cal N}_{\varsigma}({\cal N}_{\varsigma}+1)
- C_{\varsigma}^2}~.
\eea
In the limit of the early universe ($\Delta' \ll \omega$),
Eqs.~(\ref{3}) reduce to
\bea 
\label{gold}
C_{\varsigma}'&=&0~,\nonumber\\
\frac{d{\cal N}_{\varsigma}}{d \Delta_{\varsigma}}&=&
-\sqrt{ 4 {\cal N}_{\varsigma}({\cal N}_{\varsigma}+1) - C_{\varsigma}^2}~.
\eea
A general solution of these equations is
\bea \label{gen}
2{\cal N}_{\varsigma}+1\equiv \cosh(2r_{\varsigma})&=&
\cosh(2\Delta_{\varsigma})
+\frac{C_{\varsigma}^2}{2}e^{-2\Delta_{\varsigma}}~,\nonumber\\
C_{\varsigma}&=&{\rm const}~.
\eea
From equations ~(\ref{gen}) it follows that
the {\it vacuum} initial state~(\ref{vacuum})
 ${\cal N}_{\varsigma}(\eta=0)=0 $ entails that
$C_{\varsigma}(\eta=0)=0$.
In terms of $r$ and $\theta$ this corresponds to the initial values 
for the cosmic evolution
\be \label{id}
r_{\varsigma}(\eta=0)=0,~~~~~~~~~~~~~~~~
{\theta}_{\varsigma}(\eta=0)=\frac{\pi}{4}.
\ee
The solution~(\ref{gen}) to eq.~(\ref{gold}) for $\omega=0$ can be treated
as the Goldstone mode
that rejects the symmetry breaking with respect to translations in time.
The numbers of created particles for the large Hubble limit
are determined by the vacuum solutions
\be \label{vacuum1}
r_{\varsigma}=\Delta_{\varsigma},
~~~~~~~~~~~~~~~{\theta}_{\varsigma}=\frac{\pi}{4}~.
\ee
In accordance with equations (\ref{Nb}) and~(\ref{Nf}), we have
\be \label{Nt}
{\cal N}_{\varsigma}^{(B)} =
\sinh^2\Delta_{\varsigma}(\vh),
~~~~~~~~~~~~~~~~~~~
{\cal N}_{\varsigma}^{(F)}=\sin^2 \Delta_{\varsigma}(\vh)~.
\ee
In particular, the number of created  fermions is equal to
\be \label{Ns}
{\cal N}_{s,k,\sigma}=
\frac{1}{2}\left(1-\frac{k^2+y_s^2\vh\vh_I}{\omega_s(\vh)\omega_s(\vh_I)}
\right)~.
\ee
It is easy to see that the relativistic limit of large momenta
prevents the cosmological creation of all particle species
except for gravitons and longitudinal bosons in accordance with
their mass singularity~\cite{hp,sf}.

\section{Early Universe Scenario}

\subsection{Rigid state}

From the action (\ref{Cw}) considered in this work one can see that at
the beginning of time $\eta \sim 0$, the dilaton goes to zero
$\vh \sim 0$  together with the potential energy, whereas the
kinetic energy of gravitons goes to infinity as $\sim 1/\vh^2$,
${\cal N}_h \sim 1/\vh^2$. This behavior is well known~\cite{M}  
from anisotropic homogeneous excitations of the metrics in a universe
with the rigid equation of state (the Kasner stage).
This corresponds to the behavior of the conformal density~(\ref{red})
\be \label{rigid}
\Omega(z)=\Omega_{\rm Rigid} (z+1)^2~,~~~~~~~~~~
\Omega_{\rm Rigid} \leq 1~,
\ee
and the  evolution of the universe~(\ref{vd})
\be\label{vda}
H(z):=\frac{\vh'}{\vh}=-\frac{z'}{z+1}= (z+1)^2 \sqrt{\Omega_{\rm Rigid}} H_0~.
\ee
The solution of this equation takes the form
\be  \label{etai}
\vh^2(\eta)=\frac{\vh_0^2}{(z+1)^2}=\vh_I^2\left[{\frac{\eta+\eta_I}{\eta_I}}\right]~,
\ee
where the subscript ``I'' denotes values at the initial time $\eta_I$. 
For definiteness, we list the values of the initial Hubble parameter $H_I$,
initial time $\eta_I$, initial z-factor $(z_I+1)$, and initial vector
boson mass $m_v(z_I)$
\be  \label{etaii}
H_I= H(z_I)=\frac{1}{2\eta_I}={(z_I+1)^2 H_0\sqrt{\Omega_{\rm Rigid}}}~,
~~~~ z_I+1=\frac{\vh_0}{\vh_I}~,~~~m_v(z_I)=\frac{m_v}{z_I+1}~.
\ee
To isolate the point of singularity,
we shall consider the beginning of time $\eta=0$ in Eq.~(\ref{etai}) with
the initial value  $\vh_I$ for dilaton as the beginning  of the
creation of the primordial vector bosons.

In the anisotropic era, the Bogoliubov equations~(\ref{main})  for
the numbers of created vector bosons (longitudinal $(||)$, and
transversal $(\bot)$) can be rewritten in terms of
dimensionless variables for time and momentum
\be \label{dless}
\tau=\eta/\eta_I,~~~~~~x=\frac{q}{m_v(z_I)}~,
\ee
and the vector boson disperson relation 
\be 
\label{omeg}
\omega_v=H_I\gamma_v\sqrt{1+\tau+x^2}~,
\ee
where $\gamma_v=\frac{m_v(z_I)}{H_I}$ 
is the vector boson mass parameter.
The corresponding Bogoliubov equations~(\ref{main})  for
the number of created bosons in the anisotropic (rigid state) era 
are defined by the functions
\bea 
\label{dt}
\Delta_{\bot}'&=&
\frac{1}{2}\frac{\omega'_v}{\omega_v}={H_I}
\frac{1}{2(1+\tau+x^2)}~,\\
\label{dl}
\Delta_{||}'&=& \frac{\vh'}{\vh}-\Delta_{\bot}'=
H_I\left[
\frac{1}{1+\tau}-\frac{1}{2(1+\tau+x^2)}\right]~.~
\eea
The initial values are given by~(\ref{id}), see also Appendix A.
The form of these equations shows us that the parameter value
$\gamma_v = 1$ is distinguished. For $\Omega_{\rm Rigid}=1$ this 
value corresponds to
$z_I + 1 = (m_v/H_0)^{1/3}\sim 3.4\times 10^{14}$.
At this high redshift $H_I=m_v(z_I)=k_B~2.76$ K, i.e. the 
dilaton field changed so rapidly as to create vector mesons which were  
light enough in that era to form upon annihilation the CMB 
the temperature of which is an invariant of the cosmic evolution.
In the following we study the cosmological creation of
vector bosons for the universe in the rigid state.

\subsection{Creation of vector bosons by the anisotropic universe}

We have calculated the number of created bosons during their lifetime
$\tau_L=\eta_L/\eta_I$.
To estimate roughly this time $\eta_L$, we use
the lifetime of W-bosons in the Standard Model at this moment
$$
\eta_L+\eta_I= \frac{\sin^2 \theta_W}{m_W(z_L)\alpha_{QED}}~,
$$
where $\theta_W$ is the Weinberg angle,  $\alpha_{QED}=1/137$, and $z_L$
is the z-factor at this time.
Using Eqs.~(\ref{etai}),~(\ref{etaii})), and the
equality $\eta_I m_W(z_L)=(\gamma_v/2)(z_I+1)/(z_L+1)$, we rewrite the
previous equation in terms of the z-factor
\be \label{life}
\tau_L+1=\frac{(z_I+1)^2}{(z_L+1)^2}=
\frac{(z_L+1)}{(z_I+1)} \frac{2\sin^2\theta_W}{\gamma_v\alpha_{QED}}~,
\ee
The solution of this equation~(\ref{life}) is
\be \label{lifes}
\tau_L+1=
\left(\frac{2\sin^2\theta_W}{\gamma_v\alpha_{QED}}\right)^{2/3}
\simeq \frac{16}{\gamma_v^{2/3}}~,
\ee
and for the timelife of created bosons we have
\be \label{lv}
\tau_L =\frac{\eta_L}{\eta_I}\simeq \frac{16}{\gamma_v^{2/3}}-1~.
\ee
In the following we consider the case $\gamma_v=1$ for which $\tau_L=15$.

The numerical solutions of the Bogoliubov equations~(\ref{main}) for
the time dependence of the vector boson distribution
functions ${\cal N}^{||}$ and ${\cal N}^{\bot}$ are given
in Fig. 1 (left panels) for the momentum $x=1.25$.
We  can see that the longitudinal function
is greater than the transversal one.

The momentum dependence of these functions at the time $\tau=14$
is given on the right panels of Fig.1.
Upper panel shows us the intensive cosmological creation of
the longitudinal bosons in comparison with
the transversal ones.
This fact is in agreement with the mass singularity of
the longitudinal vector bosons discussed in Refs.~\cite{sf,hp}.

One of the features of this intensive creation is a high momentum tail
of the momentum distribution of longitudinal bosons which leads to a
divergence of  the density of created particles defined as~\cite{par}
\be\label{nb}
n_{v}(\eta_L)=\frac{1}{2\pi^2}
\int\limits_{0 }^{\infty }
dq q^2
\left[ {\cal N}^{||}(q,\eta_L) + 2{\cal N}^{\bot}(x,\eta_L)\right]~.
\ee
The divergence is a defect of our approximation,
where we neglected all interactions of vector bosons, that form
the collision integral in the kinetic equation for the distribution
functions.

In order to obtain a finite result for the density we suggest to multiply
the primordial distributions
${\cal N}^{||}(x,\eta_L)$ and ${\cal N}^{\bot}(x,\eta_L)$ with
the Bose - Einstein distribution $(k_B=1)$
\be \label{bose}
{\cal F}\left(\frac{q}{T}, \frac{m_v(\tau)}{T},
\frac{\mu(\tau)}{T}
\right)=
\left\{\exp\left[\frac{\omega_v(\tau)-\mu(\tau)}{T}\right]
-1\right\}^{-1}~,
\ee
where $T$ is considered as the regularization parameter.
Then, the density is given as
\be\label{n1}
n_{v}(T,\eta_L)=\frac{T^3}{2\pi^2}
\int\limits_{0 }^{\infty }
dy y^2 {\cal F}\left(y,
\gamma_T,
\gamma_{\mu}
\right)
\left[ {\cal N}^{||}\left(\frac{y}{\gamma_T},\eta_L\right)
    + 2{\cal N}^{\bot}\left(\frac{y}{\gamma_T},\eta_L\right)\right]
\ee
for each vector boson $(v=W^{\pm},Z^0)$, where
\be \label{gt}
\gamma_T=\frac{m_v(z_I)}{ T}~,~~~~
\gamma_{\mu}=\frac{\mu(z_I)}{ T}~.
\ee
The problem is to find the value of this density.
\begin{figure}[ht]
\includegraphics[width=\textwidth]{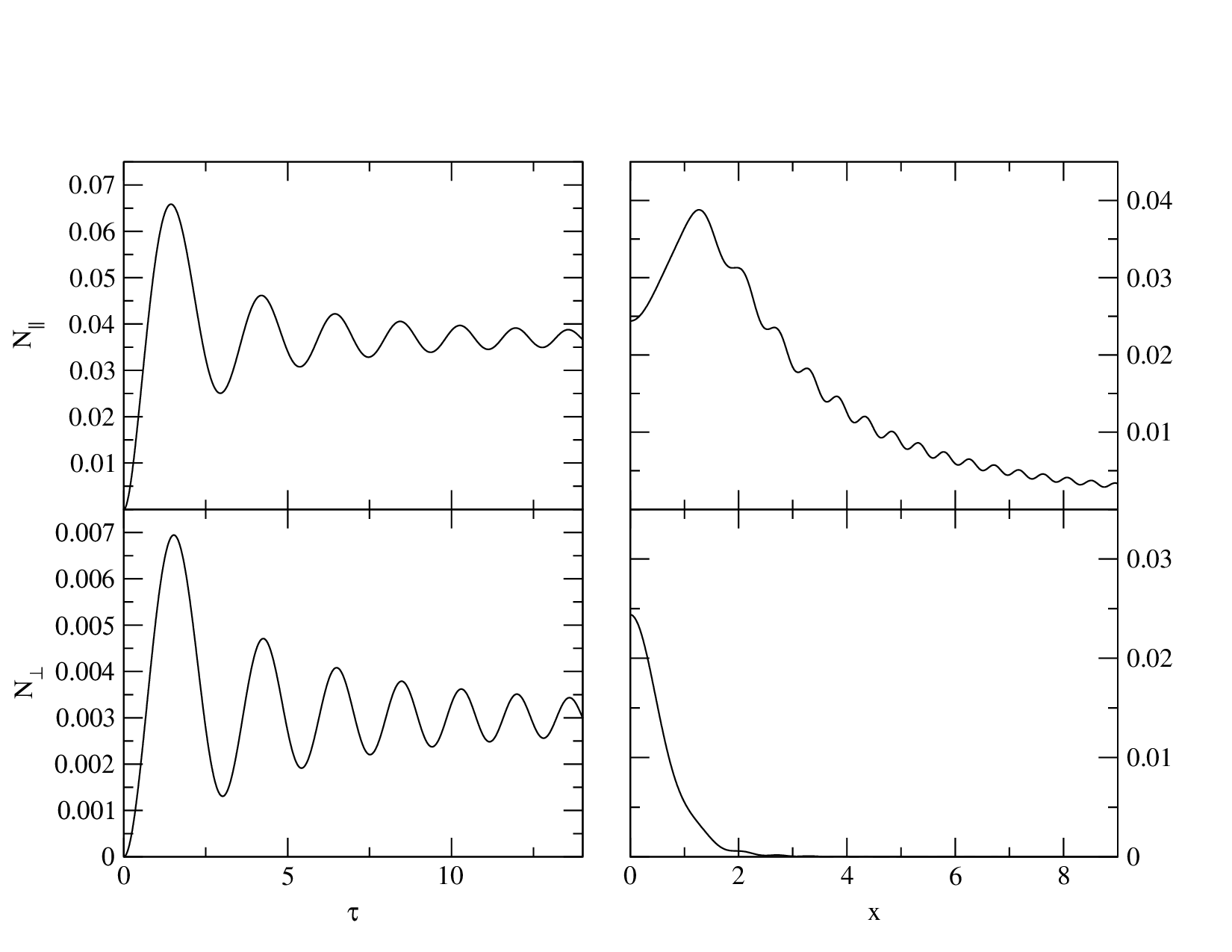}
  \caption{Time dependence  for the dimensionless momentum
 $x=1.25$ (left panels) and momentum dependence  at the dimensionless time
 $\tau=\tau_L=14$ (right panels) of the transverse (lower panels) and
 longitudinal
 (upper panels) components of the vector boson distribution function.}
 \end{figure}

Our calculation presented on Fig.1 signals that the density~(\ref{n1})
is established very quickly in comparison with
the lifetime of bosons, and in the equilibrium
there is a weak dependence of the density on the time (or z-factor).
This means that
the initial Hubble parameter $H_I$  almost
coincides with Hubble parameter at the point of the saturation $H_s$.

 For example we calculated the values of integrals (\ref{nb})
  for the regularization parameter
\be \label{tcmbr}
T= m_v(z_I)= H_I~.
\ee
This choice corresponds to $\gamma_v =\gamma_T = 1, \mu=m_v $.
The result of the calculation is
\be \label{nc}
\frac{n_{v}}{T^3} = \frac{1}{2\pi^2}
\left\{ [1.877]^{||}+2 [0.277]^{\bot}=2.432   \right\}~,
\ee
 where we denote the contributions
of the longitudinal and transverse bosons by the labels $(||,~ \bot)$.

After the time $\tau_L$ primordial bosons decay.
The final product of the decay of the primordial bosons includes photons.
If one photon goes from the annihilation of the products of decay
of $W^{\pm}$ bosons, and another photon - from $Z$ bosons, we can
expect the density of photons in the Conformal Cosmology with
the constant temperature and a static universe~\cite{12}
\be \label{1nce}
\frac{n_{\gamma}}{T^3}=\frac{1}{\pi^2}
\left\{ 2.432  \right\}.
\ee
By the comparison of this value with
the present-day density of
the cosmic microwave radiation
\be \label{1nc}
\frac{n^{\rm obs}_{\gamma}}{T_{\rm CMB}^3}=\frac{1}{\pi^2}
\left\{ 2 \zeta (3)= 2.402  \right\}.
\ee
we can estimate the regularization parameter $T$.
One can see that this parameter  is the order of the
temperature of the cosmic microwave background
\be \label{eqc}
T=T_{\rm CMB}= 2.73~{\rm K}.
\ee

We can speak about thermal equilibrium for the primordial bosons with
a temperature $T_{\rm eq}$,
if the  inverse relaxation time~\cite{ber}
\be \label{rel0}
\eta_{\rm rel}^{-1}(z_I) = {\sigma_{\rm scat }n_{v}(T_{\rm eq})}
\ee
where the scattering cross-section of bosons in the considered region
is proportional to the inverse of their squared mass
\be \label{gs}
\sigma_{\rm scat}=\frac{\gamma_{\rm scat}}{m^2_v(z)}~
\ee
is greater than the primordial Hubble parameter $H_I$.
This means that the thermal equilibrium
will be maintained
\be \label{rel01}
\gamma_{\rm scat}n_v(T_{\rm eq }) =
\frac{2.4~\gamma_{\rm scat}}{\pi^2}T_{\rm eq}^3
= H(z_I)m^2_v(z_I) = H(0)m^2_v(0).
\ee
The right hand side of this formula
is an integral of motion for the evolution of the universe
in the rigid state. The estimation of this integral
from the present values of the Hubble parameter and boson mass gives
the value
\be \label{tcmbr1}
 \left[m_W^{2}(z_I)H(z_I)\right]^{1/3}=\left[m_W^{2}(0)H_0\right]^{1/3} = 2.76~{\rm K}.
\ee
Thus we conclude that the assumption of a quickly established
thermal equilibrium in the primordial vector boson system may be
justified since $T\sim T_{\rm eq }$. The temperature of the photon
background emerging after annihilation and decay processes
of $W^{\pm}$ and $Z$ bosons is invariant in the Conformal Cosmology
and the simple estimate performed above gives a value
surprisingly close to that of the observed CMB radiation.

A more detailed investigation of the kinetic processes
which govern the transition from the primordial vector boson era
to the photon era can be based on a solution of the corresponding kinetic
equations \cite{s} and will be given elsewhere.

\subsection{The baryon asymmetry}

The baryon asymmetry of the universe appears as a result of the polarization
of the Dirac vacuum of quarks by transversal bosons in accordance with
the selection rule of the Standard Model~\cite{ru}
\be \label{SR}
\Delta L = \Delta B = \Delta n_w + \Delta n_z~,
\ee
where
\be \label{nw}
\Delta n_W = \frac{4{\alpha}_{QED}}{\sin^{2}\theta_{W}}J^W,~~~
J^W=\int_{0}^{\eta^{\rm W}_{\rm l}} d\eta
\int \frac{d^3 x}{4\pi}~~{}_{\rm sq}<0|E^{W}_{i}B^{W}_{i}|0>{}_{\rm sq}~,
\ee
\be \label{nz}
\Delta n_Z = \frac{{\alpha}_{QED}}{\sin^{2}\theta_{W}\cos^{2}\theta_{W}}J^Z,~~~
J^Z=\int_{0}^{\eta^{\rm Z}_{\rm l}} d\eta
\int  \frac{d^3 x}{4\pi}~~{}_{\rm sq}<0|E^{Z}_{i}B^{Z}_{i}|0>{}_{\rm sq}
\ee
are the topological
winding number functionals of the primordial  $W^{\pm}$ and $Z$ bosons,
and $E_i$, $B_i$ are the electric and magnetic field strengths.
The squeezed vacuum gives a nonzero value for these quantities
\be \label{eb}
\int \frac{d^3 x}{4\pi}~~{}_{\rm sq}<0|E^{v}_{i}B^{v}_{i}|0>{}_{\rm sq} =
-\frac{V_0}{2} \int\limits_{0 }^{\infty }dk |k|^3 C_v(\eta,k)~,
\ee
where $C_v(\eta,k)$ is given by the equation~(\ref{sub}) for the
transversal bosons. We estimated $J^W/V_0T^3\approx 1.44$
and $J^Z/V_0T^3\approx 2.41$ for $\gamma_v=1$
and a timelife of the bosons $\tau_L^W\approx 15$, $\tau_L^Z\approx 30$,
using the T-regularization~(\ref{bose}).

Thus, we can see that
the baryon asymmetry can be explained by the topological
winding number functional of the primordial bosons and
the superweak interaction of $d$ and $s$-quarks
$(d+s~\rightarrow ~s+d)$
with CP-violation, experimentally observed in the decays of
$K$ mesons~\cite{o}.
A more detailed consideration of the baryon asymmetry phenomenon
will be presented in a subsequent paper.

\section{Resum\'e}

We have considered the simplest Cold Universe Scenario where
the physical reason of CMB is the cosmic creation of primordial
longitudinal vector bosons.
Among the matter fields there is only one longitudinal component
of the vector bosons with a singular behavior of the integral
of motion at this region. To see this singularity, we consider only
the mass term for the time component $v_0$ that is proportional to
the time derivative of the longitudinal component
(due to the constraint $v_0\sim \dot v_{||}$).
The toy Lagrangian of the conformal universe filled in
these bosons takes the form
\be \label{toy1}
{\cal L}={\cal L}_{\vh}+{\cal L}_{||}=
- \left(\frac{d \vh}{d\eta}\right)^2+ \vh^2(\eta) \left(\frac{d v_{||}}{d\eta}\right)^2
\ee
with the equations of motion
\be \label{toy2}
\frac{d^2 (\vh)}{d \eta^2} +2\vh\left(\frac{d v_{||}}{d\eta}\right)^2=0~,
\ee
\be \label{toy1e}
\frac{d}{d\eta} \left[\vh^2(\eta) \frac{dv_{||}}{d\eta}\right]=0~.
\ee
The last equation entails the existence of an integral of motion
\be \label{toy1ei}
\left[\vh^2(\eta)\frac{d v_{||}}{d\eta}) \right]=P_{v}.
\ee
In terms of this integral of motion the
Lagrangian~(\ref{toy1}) and the energy of matter takes the
form of the rigid state
\be \label{toy1s}
{\cal L}_{||}=\frac{P_{v}^2}{\vh^2(\eta)}~,
\ee
whereas the equation of motion for the dilaton~(\ref{toy2}) becomes
trivial for the square of the dilaton
\be \label{toy21}
\frac{d^2 (\vh^2)}{d \eta^2} =0~.
\ee
A  solution of this classical equation is defined
by an initial position  $\vh_I$  and velocity  $H_I$  of the
dilaton
\be \label{toy2e}
\vh^2(\eta)=\vh_I^2(1 +2H_I \eta)
\ee
$H_I$ coincides with the initial Hubble parameter $\varphi'_I/\varphi_I$.

The numerical solution of the exact Bogoliubov equations shows us that
the temperature equilibrium is established so quickly that
the z-factor and the Hubble parameter almost do not change.
The latter determine the temperature of the primordial bosons as
the integral of motion
\be\label{ht2}
T_{\rm eq}=[H_0 m_W^2(0)]^{1/3}=2.76~ {\rm K}~,
\ee
which almost coincides
with the present-day value of the cosmic microwave radiation.

Thus, in the context of the relative standard of measurements
and the Conformal Cosmology, the estimations of temperature
and density of the primordial vector bosons show us that
the origin of the cosmic microwave radiation and the observable
matter in the  universe can be seen in the decay of the primordial vector
bosons into photons, leptons, and  quarks.
These primordial vector bosons are created by the conformal universe
at the time $\eta_I=10^{-12}$ that corresponds to the initial z-factor
\be \label{zi}
z_I+1=\frac{m_W(0)^{1/3}}{H_0^{1/3}} \simeq 3.5 \times 10^{14}~.
\ee

It is worth to emphasize that in the considered model of the
Conformal Cosmology \cite{12}, the temperature~(\ref{ht2})  is a constant.
In Conformal Cosmology, we have the mass history
\be
\label{mz}
m_{\rm era}{(z)}=\frac{m_{\rm era}(0)}{(1+z)}=T_{\rm eq}
\ee
with the constant temperature $T_{\rm eq}= 2.73 ~ {\rm K} = 2.35
 \times 10^{-13}$ GeV,
where $m_{\rm era}(0) $ is a characteristic energy (mass)
of an era of the universe evolution.

Eq. (\ref{mz}) has the important consequence that all those  physical
processes which concern the chemical composition of the universe
and  which depend basically on Boltzmann factors with the argument $(m/T)$
cannot distinguish between the Conformal Cosmology an the Standard
Cosmology due to the relations
\be\nonumber
\frac{m(z)}{T(0)}=\frac{m(0)}{(1+z)T(0)}=\frac{m(0)}{T(z)}~.
\ee
This formula makes transparent that in this order of approximation
a $z$-history of masses with invariant temperatures in the rigid state
of Conformal Cosmology is equivalent to
a $z$-history of temperatures with invariant masses in the radiation stage
of the Standard Cosmology.
We expect therefore that the Conformal Cosmology allows us to keep the 
scenarios developed in the Standard Cosmology in the radiation stage for, 
e.g. the neutron-proton ratio and primordial element abundances.

Recall that the theoretical foundation of the radiation stage in the 
Standard Cosmology became problematic in the light of the new Supernova data
on the present-day accelerating evolution of the universe at the dominance of 
the inflation stage. In this situation it is difficult to explain in the 
Standard Cosmology the status of the radiation stage that follows the 
primordial inflation stage and is followed by a short matter era which in turn
is terminated by the present-day inflation stage.
On the other hand, in the Conformal Cosmology a unique permanent rigid state 
can explain the primordial creation of matter from the vacuum, the primordial 
element abundances, and the recent acceleration witnessed by distant type Ia
supernovae \cite{ps,12}.   

An important new feature of the Conformal Cosmology relative to the
Standard one is the absence of the Planck era, since
the Planck mass is not a fundamental parameter but only
the present-day value of the  dilaton field~\cite{12,plb}.

\section{Conclusion}

We have considered a unified theory of all interactions in a
space-time with the Weyl relative standard of
measurement. 
The laws of nature, i.e. the equations of motion, in this theory 
are conformal invariant and do not contain any dimensional parameter,
whereas the initial data violate the conformal symmetry.
We have shown that this unified theory leads to a conformal version of the 
Standard Cosmology which is free of those problems related to the expansion 
of the universe. 
In the Conformal Cosmology, instead of a z-history of the temperature we have 
obtained a z-history of masses at constant temperature where the number of 
created particles is determined by the initial data of the universe.

We have shown that the conformal symmetry, the reparametrization-invariant 
perturbation theory and the mass-singularity of the lomgitudinal components of 
vector bosons lead to the effect of an intensive creation of these bosons with 
the temperature 
$(m_W^2H_0)^{1/3} \sim  2.73 ~ {\rm K}$ and a density which
resembles physical properties of the cosmic microwave background
radiation.
We have derived a similar enhancement of particle creation
in the case of the Higgs field, but we
do not consider this case in detail because
its existence is not experimentally proven yet.

According to the scenario outlined in this work, the primordial boson 
radiation is created during a conformal time
interval of $2 \times 10^{-12}~{\rm sec}$ and subsequent annihilation
and decay has formed all the matter in the universe.

The further z-history of the conformal universe
repeats that of the Friedmann universe,
with a remarkable difference :
instead of the z-dependence of the temperature
in an {\it expanding universe} with constant masses (Standard Cosmology),
we have a z-history of masses in a {\it static universe}
with almost constant temperature of the photon background in Conformal 
Cosmology.

The  Conformal Cosmology
 provides  definite solutions to the problems of Standard Cosmology.
The Minkowski flat space has no horizon problem. The problems
of the Planck age  also do not exist as the Planck mass
is not a fundamental parameter of the theory but
only the ordinary present-day value of the dilaton field.

In the present paper we have added to this appealing concept of
Conformal Cosmology a physical mechanism explaining the origin 
of the matter content of the
universe by pair creation of longitudinal vector bosons from the
dilaton field which, in particular, gives a surprisingly good
estimate for the temperature of the CMB radiation.

\vspace{4mm}

{\bf Acknowledgments}

The authors are grateful to A.V.~Efremov, V.P.~Karasev,
V.V.~Kornyak, and E.A.~Kuraev for fruitful discussions.
DP,~SV, and AG thank RFBR (grant 00-02-81023 Bel 2000$\_a$) for support.

\section*{ Appendix A}

The Bogoliubov equations for the Higgs field $(\chi)$ and vector bosons ($v$)
in terms of dimensionless variables and parameters
(\ref{dless})-(\ref{dt}) take the form
$$
\left[\frac{\gamma_v}{2}\sqrt{(1+\tau)y^{2}_{\chi}+x^2} +
\frac{d}{d\tau}\theta_{\chi}\right] \tanh(2r_{\chi}) =
\frac{1}{2}\left[\frac{1}{(1+\tau)y^{2}_{\chi}}+\frac{1}{2}\frac{1}{\left[(1+\tau)y^{2}_{\chi}+x^2\right]}\right]
\cos(2\theta_{\chi})~,
$$
$$
\frac{d}{d\tau}r_{\chi} =
-
\frac{1}{2}\left[\frac{1}{(1+\tau)y^{2}_{\chi}}+\frac{1}{2}\frac{1}{\left[ (1+\tau)y^{2}_{\chi}+x^2\right]}\right]
\sin(2\theta_{\chi})~,
$$
$$
\left[\frac{\gamma_v}{2}\sqrt{(1+\tau)+x^2} -
\frac{d}{d\tau}\theta^{||}_{v}\right] \tanh(2r^{||}_{v}) =
\frac{1}{2}\left[\frac{1}{(1+\tau)}-\frac{1}{2}\frac{1}{\left[ (1+\tau)+x^2\right]}\right]
\cos(2\theta^{||}_{v})~,
$$
$$
\frac{d}{d\tau}r^{||}_{v} =
-
\frac{1}{2}\left[\frac{1}{(1+\tau)}-\frac{1}{2}\frac{1}{\left[ (1+\tau)+x^2\right]}\right]
\sin(2\theta^{||}_{v})~,
$$
$$
\left[\frac{\gamma_v}{2}\sqrt{(1+\tau)+x^2} -
\frac{d}{d\tau}\theta^{\bot}_{v}\right]  \tanh(2r^{\bot}_{v}) =
\frac{1}{4}\left[\frac{1}{(1+\tau)+x^2}\right]
\cos(2\theta^{\bot}_{v})~,
$$
$$
\frac{d}{d\tau}r^{\bot}_{v} =
-\frac{1}{4}\left[\frac{1}{(1+\tau)+x^2}\right]
\sin(2\theta^{\bot}_{v})~,
$$
where $y^{2}_{\chi}$ is defined as
$$
y^{2}_{\chi}= 4\lambda\frac{y^{2}_{h}}{y^{2}_{v}}~.
$$
The creation of t-quarks with the mass
$$
m_t=\frac{y_v}{y_s}m_W :=\gamma_s m_W~~~~~~~~(\gamma_s \simeq 2 )~,
$$
is described by $r_t,\theta_t$ with the equations
$$
\left[\frac{\gamma_v}{2}\sqrt{\gamma_s^2(1+\tau)+x^2} -
\frac{d}{d\tau}\theta_{t}\right] \sin(2r_{t}) =
\frac{1}{4}\left[\frac{1}{(1+\tau)+x^2/\gamma_s^2}\right]
\cos(2\theta_{t})\cos(2r_{t})~,
$$
$$
\frac{d}{d\tau}r_{t} =
-\frac{1}{4}\left[\frac{1}{(1+\tau)+x^2/\gamma_s^2}\right]
\sin(2\theta_{t})~,
$$
and the density
$$
\frac{<n_{t}(\eta_L)>}{T^3}=
\frac{\Omega_{\rm A}^{1/2}\gamma_v}{\pi^2} \int\limits_{0 }^{\infty }
dx x^2 {\cal F}_t(x) \sin^2r_t(\tau_L)~,
$$
where
$$
{\cal F}_t(x)=
\left[\exp\left\{\gamma_T(\sqrt{\gamma_s^2(1+\tau_L)+x^2}-
\sqrt{\gamma_s^2(1+\tau_L)})\right\}+1\right]^{-1}~.
$$



\begin{thebibliography}{}
\bibitem{N}
J.V. Narlikar, Space Sci. Rev. {\bf 50} (1989) 523.
\bibitem{ps}
A.G. Riess et al., Astron. J. {\bf 116} (1998) 1009;\\
S. Perlmutter et al., Astrophys. J. {\bf 517} (1999) 565.
\bibitem{12}
D. Behnke et al., Phys. Lett. B (2001) (submitted),
gr-qc/0102039.
\bibitem{sf}
V.I. Ogievetsky, I.V. Polubarinov, ZHETF {\bf 41} (1961)  246;\\
A.A. Slavnov, L.D. Faddeev, TMF {\bf 3} (1970) 18.
\bibitem{hp}
H.-P. Pavel,  V.N. Pervushin, Int. J. Mod. Phys. A {\bf 14} (1999) 2285.
\bibitem{bt} K.A. Bronnikov, E. A. Tagirov, Preprint JINR P2-4151,
Dubna, 1968.
\bibitem{par}
G.L. Parker,  Phys. Rev. Lett. {\bf 21} (1968) 562,
Phys. Rev. {\bf 183} (1969)  1057,
Phys. Rev. {\bf D 3} (1971)  346.
\bibitem{gmm}
A.A. Grib, S.G. Mamaev, V.M. Mostepanenko
{"Quantum effects in intensive external fields"}
(Moscow, Atomizdat, 1980) (in Russian).
\bibitem{zel} Ya.B. Zel'dovich, A.A. Starobinski, ZHETF {\bf 61} (1971) 2161.
\bibitem{plb}M. Pawlowski, V.V. Papoyan, V.N. Pervushin, V.I. Smirichinski,
Phys.~Lett.~B {\bf 444} (1998) 293.
\bibitem{grg}
L.N. Gyngazov, M. Pawlowski, V.N. Pervushin, V.I. Smirichinski,
Gen. Rel. and Grav. {\bf 30} (1998) 1749.
\bibitem{pp}
M. Pawlowski, V.N. Pervushin, Int. J. Mod. Phys. {\bf 16} (2001) 1715;
[hep-th/0006116].
\bibitem{ps1}
V.N. Pervushin, V.I. Smirichinski,
J. Phys. A: Math.~Gen. {\bf 32} (1999) 6191.
\bibitem{we}
H. Weyl, Sitzungsber.d. Berl. Akad., (1918) 465.
\bibitem{pct}
R. Penrose,
"Relativity, Groups and Topology",
(Gordon and Breach, London, 1964);\\
N. Chernikov, E. Tagirov, Ann. Ins. Henri Poincare,  {\bf 9} (1968) 109.
\bibitem{p1}
V. N. Pervushin et al.,Phys. Lett. {\bf B 365} (1996) 35.
\bibitem{L}
A. Lichnerowicz, Journ. Math. Pures and Appl. B {\bf 37} (1944) 23.
\bibitem{M}
C. Misner, Phys. Rev. {\bf 186} (1969) 1319.
\bibitem{ber}
J. Bernstein, "Kinetic theory in the expanding universe"
 (Cambridge University Press, Cambridge, 1985).
\bibitem{s}
For recent kinetic formulation of pair production in strong field, see
S.~Schmidt, D.~Blaschke, G.~R\"opke, A.V.~Prozorkevich,
S.A.~Smolyansky, V.D.~Toneev, Phys. Rev. {\bf D 59} (1999) 094005.
\bibitem{ru} A. D. Sakharov, JETP Lett. {\bf 5} (1967) 24;\\
V. A.~Matveev, V.A.~Rubakov, A.N.~Tavkhelidze, M.E.~Shaposhnikov,
Usp.~Fiz.~Nauk {\bf 156} (1988) 253.
\bibitem{o} L. B. Okun,
{"Leptons and Quarks"} (Moscow, Nauka, 1981) (in Russian).
\end{thebibliography}
\end{document}